\documentclass{article}
\usepackage{spconf,amsmath,graphicx}
\usepackage{amssymb,bm,upgreek}
\usepackage{nicefrac}
\usepackage[hidelinks]{hyperref}
\usepackage{multicol}
\usepackage{subcaption}
\usepackage{xcolor}
\usepackage{flushend}

\title{A Novel Method for Obtaining Diffuse field Measurements for Microphone Calibration}
\name{{Noman Akbar$^*$, Glenn Dickins$^{*\ddag}$, Mark R. P. Thomas$^\dagger$, Prasanga Samarasinghe$^*$, Thushara~Abhayapala$^*$} \thanks{This research work is funded by Australian Research Council Linkage Project LP160100379.}}
\address{$^*$Australian National University, $^\ddag$Audinate Sydney, $^\dagger$Dolby Laboratories USA \\ \{noman.akbar, prasanga.samarasinghe, thushara.abhayapala\}@anu.edu.au \\g@dickins.com, mark.r.thomas@ieee.org}

\begin{document}
\ninept
\maketitle

\begin{abstract}
We propose a straightforward and cost-effective method to perform diffuse soundfield measurements for calibrating the magnitude response of a microphone array. Typically, such calibration is performed in a diffuse soundfield created in reverberation chambers, an expensive and time-consuming process. A method is proposed for obtaining diffuse field measurements in untreated environments. First, a closed-form expression for the spatial correlation of a wideband signal in a diffuse field is derived. Next, we describe a practical procedure for obtaining the diffuse field response of a microphone array in the presence of a non-diffuse soundfield by the introduction of random perturbations in the microphone location. Experimental spatial correlation data obtained is compared with the theoretical model, confirming that it is possible to obtain diffuse field measurements in untreated environments with relatively few loudspeakers. A $30$ second test signal played from $4-8$ loudspeakers is shown to be sufficient in obtaining a diffuse field measurement using the proposed method. An Eigenmike\textsuperscript{\textregistered} is then successfully calibrated at two different geographical locations.
\end{abstract}
\begin{keywords}
Spherical arrays, microphone calibration, diffuse field measurements, and spatial correlation.
\end{keywords}
\section{Introduction} \label{sec:intro}
Diffuse field measurements are important for calibrating and testing the magnitude response of acoustic devices. Reverberation chambers are often used to create a diffuse field \cite{Bradley2015, Morrow1971, Chun2003, Cook1955}. However, they are expensive and may have limited availability. Alternatively, a diffuse field measurement can be derived from a dense set of anechoic measurements \cite{Politis2017}. Such measurements are also difficult to obtain as anechoic chambers and robotic loudspeaker mounts are expensive and not very common. A practical solution for obtaining a diffuse field response in regular rooms is therefore desirable.

A diffuse field is defined as an acoustic field where the energy density is uniform in all directions \cite{Cook1955,Nelisse1997}. Spatial correlation is an important metric for characterizing diffuse fields \cite{Chun2003,Teal2002,Rafaely2000,Kuster2007}. While acoustic signals are wideband in nature, most existing research works analyzes spatial correlation of narrowband signals in a diffuse field, which may be restrictive in several practical scenarios. Instead, we consider the spatial correlation of wideband signals in diffuse fields.

We propose a novel method for obtaining the diffuse field response of a microphone array in the presence of a non-diffuse soundfield. The main contributions of this work are:
\begin{enumerate}
\item Deriving a closed-form expression for correlation of wideband signals in diffuse fields.
\item Proposing a method to produce diffuse field measurements in a non-diffuse soundfield by perturbing the microphone array location during data capture. Experimental results using the proposed method are shown to agree with the theoretical results.
\item Using the proposed method for diffuse field measurement to successfully calibrate an Eigenmike\textsuperscript{\textregistered} at two different geographical locations demonstrating the efficiency and reproducibility of the proposed method.
\end{enumerate}

\section{Spatial Correlation of Wideband Signals in a Diffuse field}
Consider an array with $M$ microphones placed at an arbitrary location in 3D space. The $p$-th and $q$-th microphones, denoted by $m_p$ and $m_q$, are placed at location $\mathbf{x}_p$ and $\mathbf{x}_q$ in the 3D space. The spatial correlation $\rho\left(\mathbf{x}_p,\mathbf{x}_q\right)$ between the signals captured by the two microphones is defined as \cite{Cook1955,Teal2002}
\begin{align} \label{corr}
\rho\left(\mathbf{x}_p,\mathbf{x}_q\right) =
\frac{\mathbb{E}\left\{s_p\left(t\right)s_q^{*}\left(t\right)\right\}}{\mathbb{E}\left\{s_p\left(t\right)s_p^{*}\left(t\right)\right\}},
\end{align}
where $s_p\left(t\right)$ and $s_q\left(t\right)$ are the signals received at $m_p$ and $m_q$, respectively. Furthermore, we assume that $s_p\left(t\right)$ is a wideband signal represented by
\begin{align} \label{sig1}
s_p\left(t\right) = \frac{1}{\Delta\omega}\int_{\omega_\textrm{min}}^{\omega_\textrm{max}}\int_{R} A\left(\widehat{\mathbf{y}}\right) e^{-i\left(\frac{\omega}{c}\right)\mathbf{x}_p\bullet\widehat{\mathbf{y}}} e^{i\omega t} d\widehat{\mathbf{y}}d\omega,
\end{align}
where $\bullet$ is the dot product between two vectors, $\Delta\omega = \omega_{\textrm{max}} - \omega_{\textrm{min}}$ specifies the bandwidth of the signal, $R$ represents a unit sphere, $c$ is the speed of sound in air, $\widehat{\mathbf{y}}$ is the unit vector in the direction of signal propagation, and $A\left(\widehat{\mathbf{y}}\right)$ is the gain of the signal arriving from the direction of $\widehat{\mathbf{y}}$ and is assumed to be frequency independent. Using \eqref{sig1} and \eqref{corr}, we obtain
\begin{align} \label{corr_sig4}
\rho\left(\mathbf{x}_p,\mathbf{x}_q\right)
&=\frac{\int_{\omega_\textrm{min}}^{\omega_\textrm{max}}\int_{R} \left|\mathbb{E}\left\{A\left(\widehat{\mathbf{y}}\right)\right\}\right|^2 e^{i\left(\frac{\omega}{c}\right)\left(\mathbf{x}_q - \mathbf{x}_p \right) \textrm{\textbullet}\widehat{\mathbf{y}}} d\widehat{\mathbf{y}}d\omega} {\int_{\omega_\textrm{min}}^{\omega_\textrm{max}}\int_{R} \left|\mathbb{E}\left\{A\left(\widehat{\mathbf{y}}\right)\right\}\right|^2 d\widehat{\mathbf{y}}d\omega},
\end{align}
which can be rewritten as
\begin{align} \label{corr_sig6}
\rho\left(\mathbf{x}_p,\mathbf{x}_q\right) &= \int_{\omega_\textrm{min}}^{\omega_\textrm{max}}\int_{R} \xi\left(\widehat{\mathbf{y}},\omega\right) e^{i\left(\frac{\omega}{c}\right)\left(\mathbf{x}_q - \mathbf{x}_p \right) \bullet \widehat{\mathbf{y}}} d\widehat{\mathbf{y}}d\omega, \notag \\
&= \frac{1}{\Delta\omega}\int_{\omega_\textrm{min}}^{\omega_\textrm{max}}\int_{R} G\left(\widehat{\mathbf{y}}\right) e^{i\left(\frac{\omega}{c}\right)\left(\mathbf{x}_q - \mathbf{x}_p \right) \bullet \widehat{\mathbf{y}}} d\widehat{\mathbf{y}}d\omega,
\end{align}
where
\begin{align} \label{norm_gain}
\xi\left(\widehat{\mathbf{y}},\omega\right) &= \frac{\left|\mathbb{E}\left\{A\left(\widehat{\mathbf{y}}\right)\right\}\right|^2 }{\int_{\omega_\textrm{min}}^{\omega_\textrm{max}}\int_{R} \left|\mathbb{E}\left\{A\left(\widehat{\mathbf{y}}\right)\right\}\right|^2 d\widehat{\mathbf{y}}d\omega}, \notag \\ &= \frac{1}{\Delta\omega}\left(\frac{\left|\mathbb{E}\left\{A\left(\widehat{\mathbf{y}}\right)\right\}\right|^2 }{\int_{R} \left|\mathbb{E}\left\{A\left(\widehat{\mathbf{y}}\right)\right\}\right|^2 d\widehat{\mathbf{y}}}\right) = \frac{G\left(\widehat{\mathbf{y}}\right)}{\Delta\omega},
\end{align}
and $G\left(\widehat{\mathbf{y}}\right)$ represents the average power gain of the signal received from an arbitrary direction $\widehat{\mathbf{y}}$.
We highlight that \eqref{corr_sig6} is a generalized expression valid for any microphone pair and bandwidth of the signal. The exponential term in \eqref{corr_sig6} can be rewritten by using the spherical harmonic expansion as \cite{Colton1992}
\begin{align}\label{pw_exp}
e^{i\left(\frac{\omega}{c}\right)\left(\mathbf{x}_q - \mathbf{x}_p \right) \bullet \widehat{\mathbf{y}}} &= 4\pi \sum_{n=0}^{\infty} i^n j_n\left(\frac{\omega}{c}\|\mathbf{x}_q-\mathbf{x}_p\|\right) \notag \\& \times \sum_{m=-n}^{n}Y_{nm}\left(\frac{\mathbf{x}_q-\mathbf{x}_p}{\|\mathbf{x}_q-\mathbf{x}_p\|}\right)Y_{nm}^{*}\left(\widehat{\mathbf{y}}\right),
\end{align}
where $j_n\left(.\right)$ is the spherical Bessel function, $Y_{nm}\left(.\right)$ represents the spherical harmonics, $n$ is the order and $m$ is the degree of the spherical harmonics. Substituting the value from \eqref{pw_exp} in \eqref{corr_sig6}, we obtain
\begin{align} \label{corr_sig7}
\rho\left(\mathbf{x}_p,\mathbf{x}_q\right) &= \frac{1}{\Delta\omega}\int_{\omega_\textrm{min}}^{\omega_\textrm{max}}\int_{R} G\left(\widehat{\mathbf{y}}\right) 4\pi \sum_{n=0}^{\infty} i^n j_n\left(\frac{\omega}{c}\|\mathbf{x}_q-\mathbf{x}_p\|\right) \notag \\
&\times \sum_{m=-n}^{n}Y_{nm}\left(\frac{\mathbf{x}_q-\mathbf{x}_p}{\|\mathbf{x}_q-\mathbf{x}_p\|}\right)Y_{nm}^{*} \left(\widehat{\mathbf{y}}\right) d\widehat{\mathbf{y}}d\omega.
\end{align}
We next simplify \eqref{corr_sig7} as
\begin{align} \label{corr_sig71}
\rho\left(\mathbf{x}_p,\mathbf{x}_q\right) &= \frac{4\pi}{\Delta\omega} \int_{\omega_\textrm{min}}^{\omega_\textrm{max}} \sum_{n=0}^{\infty} i^n j_n\left(\frac{\omega}{c}\|\mathbf{x}_q-\mathbf{x}_p\|\right)  \notag \\ & \times \sum_{m=-n}^{n} Y_{nm}\left(\frac{\mathbf{x}_q-\mathbf{x}_p}{\|\mathbf{x}_q-\mathbf{x}_p\|}\right)  \beta_{nm} d\omega,
\end{align}
where
\begin{align}\label{beta}
\beta_{nm} = \int_{R} G\left(\widehat{\mathbf{y}}\right) Y_{nm}^{*}\left(\widehat{\mathbf{y}}\right) d\widehat{\mathbf{y}}.
\end{align}
Assuming that the plane wave is received by the microphones $m_p$ and $m_q$ from all directions, i.e., the field is diffuse, and $\beta_{00}$=1, expression \eqref{corr_sig71} is further simplified as
\begin{align} \label{corr_sig10}
\rho\left(\mathbf{x}_p,\mathbf{x}_q\right) &= \frac{1}{\Delta\omega}\int_{\omega_\textrm{min}}^{\omega_\textrm{max}} j_0\left(\frac{\omega}{c}\|\mathbf{x}_q-\mathbf{x}_p\|\right) d\omega, \notag \\
& = \frac{1}{\Delta\omega}\int_{\omega_\textrm{min}}^{\omega_\textrm{max}} \frac{\sin\left(\frac{\omega}{c}\|\mathbf{x}_q-\mathbf{x}_p\|\right)}{\left(\frac{\omega}{c}\|\mathbf{x}_q-\mathbf{x}_p\|\right)}d\omega.
\end{align}
Defining $\omega_c = \nicefrac{\omega_{\textrm{min}} + \omega_{\textrm{max}}}{2}$, \eqref{corr_sig10} can be rewritten as
\begin{align} \label{corr_sig11}
\rho\left(\mathbf{x}_p,\mathbf{x}_q\right)& = \frac{1}{\Delta\omega}\int_{-\frac{\Delta\omega}{2}}^{\frac{\Delta\omega}{2}} \frac{\sin\left(\frac{\omega+\omega_c}{c}\|\mathbf{x}_q-\mathbf{x}_p\|\right)}{\frac{\omega+\omega_c}{c}\|\mathbf{x}_q-\mathbf{x}_p\|}d\omega,
\end{align}
which is expanded by using the identity $\sin\left(a+b\right) = \sin\left(a\right)\cos\left(b\right) + \cos\left(a\right)\sin\left(b\right)$ to obtain
\begin{align} \label{corr_sig12}
& \rho\left(\mathbf{x}_p,\mathbf{x}_q\right) = \frac{1}{\Delta\omega}\int_{-\frac{\Delta\omega}{2}}^{\frac{\Delta\omega}{2}} \frac{\sin\left(\frac{\omega}{c}\|\mathbf{x}_q-\mathbf{x}_p\|\right)
\cos\left(\frac{\omega_c}{c}\|\mathbf{x}_q-\mathbf{x}_p\|\right)}{\frac{\omega+\omega_c}{c}\|\mathbf{x}_q-\mathbf{x}_p\|} \notag \\
&+ \frac{\cos\left(\frac{\omega}{c}\|\mathbf{x}_q-\mathbf{x}_p\|\right) \sin\left(\frac{\omega_c}{c}\|\mathbf{x}_q-\mathbf{x}_p\|\right)}{\frac{\omega+\omega_c}{c}\|\mathbf{x}_q-\mathbf{x}_p\|} d\omega.
\end{align}
Next, the integral in \eqref{corr_sig12} is evaluated. Using the Taylor series expansion of the term $\frac{\omega+\omega_c}{c}\|\mathbf{x}_q-\mathbf{x}_p\|$ and neglecting the higher powers of $\frac{\Delta\omega}{\omega}$, \eqref{corr_sig12} is simplified as
\begin{align} \label{corr_sig13}
&\rho\left(\mathbf{x}_p,\mathbf{x}_q\right) = \frac{\sin\left(\frac{\Delta\omega}{2c}\|\mathbf{x}_q-\mathbf{x}_p\|\right)}{\frac{\Delta\omega}{2c}\|\mathbf{x}_q-\mathbf{x}_p\|} \Bigg[\frac{\sin\left(\frac{\omega_c}{c}\|\mathbf{x}_q-\mathbf{x}_p\|\right)}{\frac{\omega_c}{c}\|\mathbf{x}_q-\mathbf{x}_p\|} \notag \\
&- 2\frac{\cos\left(\frac{\omega_c}{c}\|\mathbf{x}_q-\mathbf{x}_p\|\right)}{\left(\frac{\omega_c}{c}\|\mathbf{x}_q-\mathbf{x}_p\|\right)^2}
\sin^2\left(\frac{\Delta\omega\|\mathbf{x}_q-\mathbf{x}_p\|}{4c}\right)\Bigg].
\end{align}
The second term in \eqref{corr_sig13} can be ignored as it decreases proportional to $\|\mathbf{x}_q-\mathbf{x}_p\|^3$ as $\|\mathbf{x}_q-\mathbf{x}_p\|$ increases \cite{Nelisse1997}. Accordingly, we get
\begin{align} \label{corr_sig14}
\rho\left(\mathbf{x}_p,\mathbf{x}_q\right)& = \frac{\sin\left(\frac{\Delta\omega}{2c}\|\mathbf{x}_q-\mathbf{x}_p\|\right)}{\frac{\Delta\omega}{2c}\|\mathbf{x}_q-\mathbf{x}_p\|} \Bigg[\frac{\sin\left(\frac{\omega_c}{c}\|\mathbf{x}_q-\mathbf{x}_p\|\right)}{\frac{\omega_c}{c}\|\mathbf{x}_q-\mathbf{x}_p\|}\Bigg],
\end{align}
which is further simplified to
\begin{align} \label{corr_sig15}
\rho\left(\mathbf{x}_p,\mathbf{x}_q\right)& = \textrm{sinc}\left(\frac{\Delta\omega}{2c}\|\mathbf{x}_q-\mathbf{x}_p\|\right) \textrm{sinc}\left(\frac{\omega_c}{c}\|\mathbf{x}_q-\mathbf{x}_p\|\right).
\end{align}
The generalized spatial correlation expression \eqref{corr_sig15} is valid for wideband signals. Note that narrowband signals are a special case of the expression \eqref{corr_sig15}, where $\Delta\omega \approx 0$ and $\omega_c = \omega$. Thus, utilizing $\textrm{sinc}(0) = 1$, the spatial correlation function of a narrowband signal is obtained from \eqref{corr_sig15} as
\begin{align} \label{corr_sig16}
\rho_{\textrm{NB}}\left(\mathbf{x}_p,\mathbf{x}_q\right)& = \textrm{sinc}\left(\frac{\omega}{c}\|\mathbf{x}_q-\mathbf{x}_p\|\right).
\end{align}

\section{Diffuse field measurement with finite loudspeaker array}
Consider a compact loudspeaker array placed in an acoustically untreated room where each speaker produces uncorrelated bandpass white noise. It is natural to ask how many loudspeakers are required to produce a diffuse soundfield such that measured spatial correlation~\eqref{corr} agrees with the theoretical result~\eqref{corr_sig15}. A novel method is proposed whereby the location of the microphone array is randomly perturbed and rotated within the loudspeaker array with a view to producing a response closer to that of a highly diffuse soundfield\footnote{A demo of the random motion for the proposed method can be viewed at \href{https://vimeo.com/367536785}{https://vimeo.com/367536785} and \href{https://vimeo.com/367536766}{https://vimeo.com/367536766}.} .

\begin{figure}
  \centering
  \includegraphics[width=20pc]{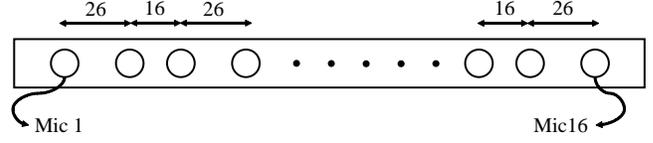}
  \caption{Linear microphone array with 16 microphones used in the experiments. The microphone spacing is in millimeters.}\label{lin_mic_array}
\end{figure}
\begin{figure*}[!t]
\captionsetup[subfigure]{justification=centering}
\begin{subfigure}{.5\textwidth}
    \centering
    \includegraphics[width=18pc]{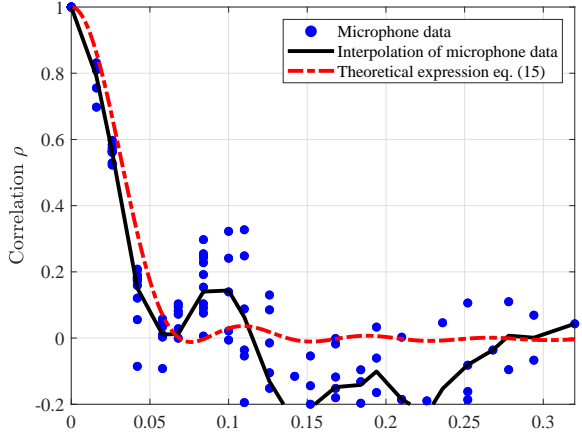}
    \caption{Experiment 1: Fixed mic array with $2$ loudspeakers playing test signal.} \quad
    \label{fig:sub-first}
\end{subfigure}
\begin{subfigure}{.5\textwidth}
    \centering
    \includegraphics[width=18pc]{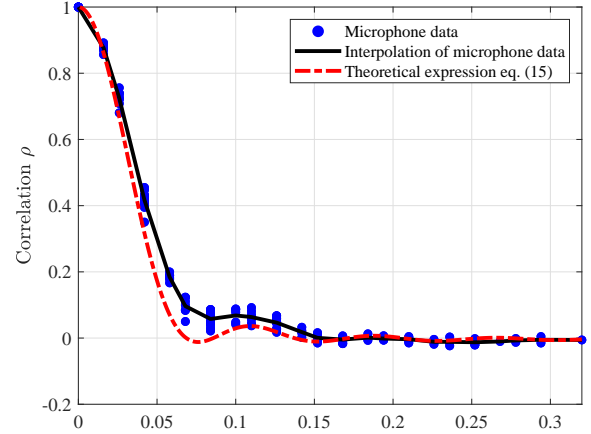}
    \caption{Experiment 2: Proposed method with $2$ loudspeakers playing test signal.} \quad
    \label{fig:sub-second}
\end{subfigure}
\begin{subfigure}{.5\textwidth}
    \centering
    \includegraphics[width=18pc]{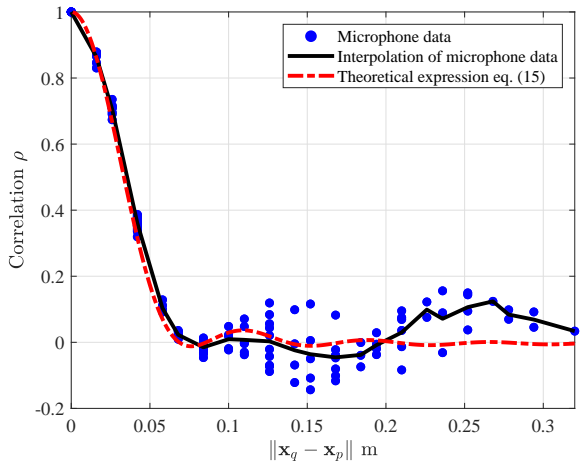}
    \caption{Experiment 1: Fixed mic array with $26$ loudspeakers playing test signal.} \quad
    \label{fig:sub-third}
\end{subfigure}
\begin{subfigure}{.5\textwidth}
    \centering
    \includegraphics[width=18pc]{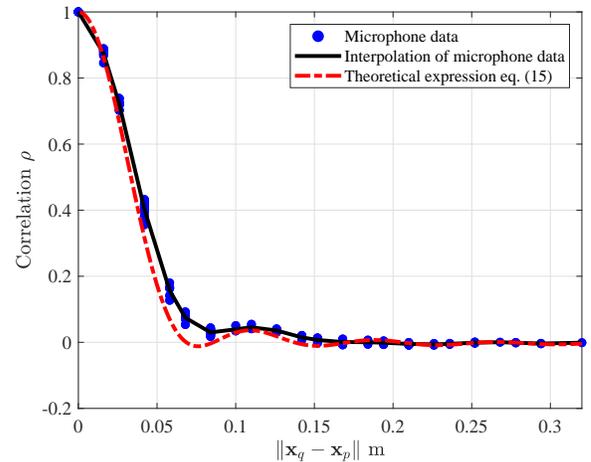}
    \caption{Experiment 2: Proposed method with $26$ loudspeakers playing test signal.} \quad
    \label{fig:sub-four}
\end{subfigure}
\caption{Microphone distance versus spatial correlation obtained from microphone data and theoretical expression for 2 and 26 loudspeakers.}
\label{corr_plots}
\end{figure*}

\subsection{Experimental Setup}
The loudspeaker array consisted of $26$ loudspeakers mounted on the vertices of a rhombic tricontrahedron with radius $1.8$~m.\footnote{Additional details about the rhombic triacontrahedron loudspeaker array model ``DAARRT$26$\textunderscore$1318$'' can be found at \href{https://vimeo.com/361493511}{https://vimeo.com/361493511}.} Loudspeakers produced uncorrelated bandpass white noise with parameters $\Delta\omega = 2\pi \times (4.5 - 0.5)~\textrm{kHz}$ and $\omega_c = 2\pi \times \nicefrac{(4.5 + 0.5)~\textrm{kHz}}{2}$. A $16$-element linear microphone array with inter-element spacing shown in~Fig.~\ref{lin_mic_array} was used to produce pairs of spatial correlation estimates with spacing between $0.016$ and $0.32$ m.

\subsection{Experiment 1: Fixed Microphone Array}
The linear microphone array was placed at the center of the loudspeaker array and $30$ seconds of data was captured using $2$ and $26$ loudspeakers. Fig.~\ref{fig:sub-first} and Fig.~\ref{fig:sub-third} depict microphone distance versus spatial correlation for all microphone pairs. While the $26$-loudspeaker case follows the theoretical curve more closely than the $2$-loudspeaker case, large variances are observed in the correlation data in both cases, suggesting that 26 loudspeakers are insufficient to produce a diffuse soundfield with a fixed microphone array.

\subsection{Experiment 2 (Proposed Method): Moving Array}
The conditions of Experiment 1 were repeated, except the microphone array location was perturbed and rotated within the loudspeaker array during capture. Fig.~\ref{fig:sub-second} and Fig.~\ref{fig:sub-four} shows the spatial correlation for the proposed method with $2$ loudspeakers and $26$ loudspeakers, respectively. Comparing Fig.~\ref{fig:sub-first} and Fig.~\ref{fig:sub-third} with Fig.~\ref{fig:sub-second} and Fig.~\ref{fig:sub-four}, we clearly observe that the variance of the spatial correlation and deviation from the theoretical model is greatly reduced with the proposed method. For example, for the proposed method with microphone distance $0.15$\;m, the variance of the microphone data is reduced by $99\%$ compared to the fixed case. Importantly, even with $2$ loudspeakers, we observe a large reduction in variance. In order to verify the repeatability of the method, the experiment was repeated with a $30$-loudspeaker dodecahedron array at a different geographical locations with three different participants who received limited instructions about movement, and similar results were obtained. This demonstrates that the measurement approximates that of a diffuse soundfield using the proposed method at multiple locations. Additionally, the method appears robust to the pattern of movement since it has little impact on the results. The test signal is received at the microphones from multiple angles and directions using the proposed method. We obtain diffuse field measurements as a result.

\subsection{Experiment 3: Restricted Loudspeaker Count}
The sensitivity of diffuseness to loudspeaker count was investigated. Table.~\ref{table} depicts the sum of variances of spatial correlations for different number of loudspeakers. It is observed that the proposed method using one loudspeaker outperforms the fixed microphone array with $26$ loudspeakers. Furthermore, the proposed method achieves an approximate diffuse field measurement with as little as $4-8$ loudspeakers.

\section{Calibrating an Eigenmike\textsuperscript{\textregistered}}
Spherical microphone arrays are widely used for 3D soundfield capture \cite{Jin2014,Thomas2019,Abhayapala2002,Meyer2002,Chen2015,Abhayapala2009}. Equipment calibration is an important step in soundfield capture, reproduction, and validation of soundfield duplication for consumer device testing \cite{Dickins2015,Gamper2016,Dickins2016}. A variant of the proposed calibration method was used in \cite{Dickins2015} to achieve a broad spectral alignment within $1$\;dB. We next demonstrate the practical significance of the proposed method by calibrating an Eigenmike\textsuperscript{\textregistered} (SN37).

\begin{figure}
  \centering
  \includegraphics[width=18pc]{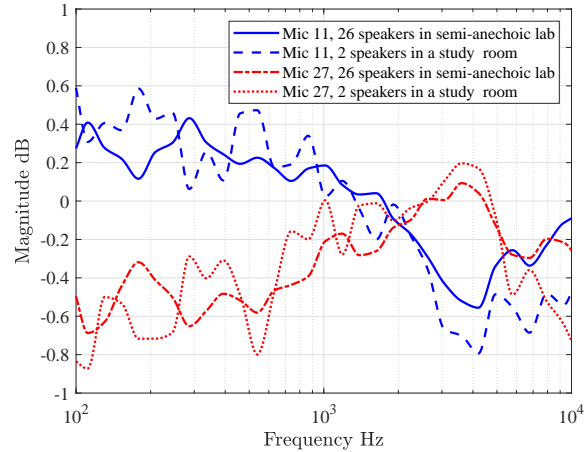}
  \caption{The Eigenmike\textsuperscript{\textregistered} magnitude response offset for Mic 11 and Mic 27 using $2$ loudspeakers and $26$ loudspeakers demonstrating repeatability of the proposed method in different environments.} \label{em32calibration}
\end{figure}

Uncorrelated pink noise was played in the $1.8$~m rhombic triacontrahedron loudspeaker array and a diffuse field measurement was obtained using the proposed method in Section.~3. We asked a participant to repeat the experiment twice, once by using $2$ loudspeakers in a normal study room and once by using $26$-loudspeaker array in a semi-anechoic room. The movement pattern of the Eigenmike\textsuperscript{\textregistered} was completely different in the two experiments. Fig.~\ref{em32calibration} depicts the consistency and repeatability of the two measurements. For this plot, the mean magnitude response of all $32$ microphones was subtracted from individual microphone magnitude responses. Importantly, the magnitude responses in the two experiments are within $0.2$\;dB. The same behaviour were observed in all the remaining $30$ microphones.

\begin{table}
  \centering
  \caption{Sum of Variances for All Microphone Distances}\label{table}
  \begin{tabular}{|l|c|c|} \hline
                & \bf{Fixed} & \bf{Proposed Method} \\ \hline \hline
   1 Speaker    & 1.2311     & 0.0135     \\ \hline
   2 Speakers   & 0.2356     & 0.0045     \\ \hline
   4 Speakers   & 0.1006     & 0.0025     \\ \hline
   8 Speakers   & 0.1418     & 0.0026     \\ \hline
   16 Speakers  & 0.0697     & 0.0017     \\ \hline
   26 Speakers  & 0.0467     & 0.0017     \\ \hline
  \end{tabular}
\end{table}
The Eigenmike\textsuperscript{\textregistered} release notes specify that the magnitude responses from all the microphones are trimmed at $1$\;kHz \cite{Acoustics2013}. Our tests indicate that the Eigenmike\textsuperscript{\textregistered} trim at $1$\;kHz has a drift of approximately $1.7$\;dB. After calibration, the magnitude responses at $1$\;kHz for all the microphones were within $0.26$\;dB. The proposed method applies calibration throughout the spectrum rather than trimming at a specific frequency. After calibration, the magnitude responses of all microphones were within $0.5$\;dB for majority of the spectrum. We calibrated the Eigenmike\textsuperscript{\textregistered} at a different geographical location and obtained similar results.

\section{Conclusion}
We proposed a method for microphone calibration by taking a diffuse field measurement with a few loudspeakers. We demonstrated the effectiveness of the proposed method by calibrating an Eigenmike\textsuperscript{\textregistered} with a $26$-speaker array in a semi-anechoic room and with $2$ speakers in a study room and achieving similar results. With the proposed method, it is possible to calibrate a microphone array such that the magnitude responses of all microphones are within $0.5$\;dB for majority of the spectrum. We demonstrated that as little as $4-8$ loudspeakers playing uncorrelated noise are sufficient to achieve a diffuse field measurement for microphone calibration. The proposed microphone calibration method is simple and cost-effective and can be performed in non-anechoic environments. The proposed method can greatly reduce the time and effort required in calibrating microphone arrays.


\begin{thebibliography}{20}
\bibliographystyle{ieeetr}

\bibitem{Bradley2015}
D.~T.~Bradley, C.~Diaz, E.~Snow, ``Improved sound field reverberance and diffusivity in a reverberation chamber through implementation of resonant-diffusing wall panels,'' \emph{Acta Acustica united with Acustic}, vol.~101, pp.~181-189, 2015.

\bibitem{Morrow1971}
C.~T.~Morrow, ``Point-to-point correlation of sound pressures in reverberation chambers,'' \emph{J. Sound and Vibration}, vol.~16, no.~1, pp.~29--42, 1971.

\bibitem{Chun2003}
I.~Chun, B.~Rafaely, and P.~Joseph, ``Experimental investigation of spatial correlation in broadband reverberant sound fields,'' \emph{J. Acoust. Soc. Amer.}, vol.~113, pp.~1995-1998, 2003.

\bibitem{Cook1955}
R.~K.~Cook, R.~V.~Waterhouse, R.~D.~Berendt, S.~Edelman, and M.~C.~Thompson, ``Measurement of correlation coefficients in reverberant sound fields,'' \emph{J. Acoust. Soc. Amer.}, vol.~27, pp.~1072--1077, 1955.

\bibitem{Politis2017}
A.~Politis and H.~Gamper, ``Comparing modeled and measurement-based spherical harmonic encoding filters for spherical microphone arrays,'' in Proc. \emph{IEEE Workshop Applications Signal Process. Audio Acoust. (WASPAA)}, New Paltz, NY, Oct. 2017, pp. 224--228.

\bibitem{Nelisse1997}
H. N\'elisse, and J. Nicolas, ``Characterization of a diffuse field in a reverberant room,'' \emph{J. Acoust. Soc. Amer.}, vol.~101, pp.~3517--3524, 1997.

\bibitem{Teal2002}
P.~D.~Teal, T.~D.~Abhayapala, and R.~A.~Kennedy, ``Spatial correlation for general distributions of scatterers,'' \emph{IEEE Signal Process. Lett.}, vol.~9, no.~10, pp.~305--308, Oct. 2002.

\bibitem{Rafaely2000}
B.~Rafaely, ``Spatial-temporal correlation of a diffuse sound field,'' \emph{J. Acoust. Soc. Amer.}, vol.~107, pp.~3254--3258, 2000.

\bibitem{Kuster2007}
M.~Kuster, ``Spatial correlation and coherence in reverberant acoustic fields: Extension to microphones with arbitrary first-order directivity,'' \emph{J. Acoust. Soc. Amer.}, vol.~123, pp.~154--162, 2007.

\bibitem{Colton1992}
D.~Colton and R.~Kress, ``Inverse acoustic and electromagnetic scattering theory''. Berlin, Germany: Springer-Verlag, 1992.

\bibitem{Jin2014}
C.~T.~Jin, N.~Epain, and A.~Parthy, ``Design, optimization and evaluation of a dual-radius spherical microphone array,'' \emph{IEEE/ACM Trans. Audio, Speech, Language Process.}, vol.~22, no.~1, pp.~193--204, Jan. 2014.

\bibitem{Thomas2019}
M.~R.~P.~Thomas, ``Practical concentric open sphere cardioid microphone array design for higher order sound field capture,'' in Proc. \emph{IEEE Intl. Conf. Acoust., Speech Signal Process. (ICASSP)}, Brighton, United Kingdom, May~2019, pp.~666--670.

\bibitem{Abhayapala2002}
T.~D.~Abhayapala and D.~B.~Ward, ``Theory and design of high order sound field microphones using spherical microphone array,'' in Proc. \emph{IEEE Intl. Conf. Acoust., Speech, Signal Process. (ICASSP)}, Orlando, FL, May~2002, pp.~1949--1952.

\bibitem{Meyer2002}
J.~Meyer and G.~Elko, ``A highly scalable spherical microphone array based on an orthonormal decomposition of the soundfield,'' in Proc. \emph{IEEE Int. Conf. Acoust., Speech, Signal Process. (ICASSP)}, Orlando, FL, May 2002, pp.~1781--1784.

\bibitem{Chen2015}
H.~Chen, T.~D.~Abhayapala, and W.~Zhang, ``Theory and design of compact hybrid microphone arrays on two-dimensional planes for three-dimensional soundfield analysis,'' \emph{J. Acoust. Soc. Amer.}, vol.~138, pp.~3081--3092, 2015.

\bibitem{Abhayapala2009}
T.~D.~Abhayapala and A.~Gupta, ``Spherical harmonic analysis of wavefields using multiple circular sensor arrays,'' \emph{IEEE Trans. Audio, Speech, Language Process.}, vol.~18, no.~6, pp.~1655-1666, 2009.

\bibitem{Dickins2015}
G.~Dickins, H.~Chen, and W.~Zhang, ``Soundfield control for consumer device testing,'' in Proc. \emph{Intl. Conf. Signal Process. Commun. Sys. (ICSPCS)}, Cairns, Australia, Dec.~2015, pp.~1--5.

\bibitem{Gamper2016}
H.~Gamper, M.~R.~P. Thomas, L.~Corbin, and I.~Tashev, ``Synthesis of device-independent noise corpora for realistic {ASR} evaluation,'' in Proc. \emph{InterSpeech}, San Francisco, CA, Aug.~2016, pp.~2791--2795.

\bibitem{Dickins2016}
G.~Dickins, P.~N.~Samarasinghe, and T.~D.~Abhayapala, ``Validation of soundfield duplication for device testing,'' in Proc. \emph{AES Conf. Sound Field Control}, Guildford, United Kingdom, Jul.~2016, pp.~1--10.

\bibitem{Acoustics2013}
mh Acoustics, ``Em32 eigenmike microphone array release notes (v17. 0),'' 25 Summit Ave, Summit, NJ 07901, USA, Oct. 2013.

\end{thebibliography}
\end{document}